# Title: Are self-citations a normal feature of knowledge accumulation?

**Authors:** Philippe Vincent-Lamarre[1], Vincent Larivière[1,2,3,4*]

**Affiliations:**
[1] École de bibliothéconomie et des sciences de l'information, Université de Montréal, Montréal, QC, Canada.
[2] School of Public Policy, Georgia Institute of Technology, Atlanta, Georgia, United States.
[3] Department of Science and Innovation-National Research Foundation Centre of Excellence in Scientometrics and Science, Technology and Innovation Policy, Stellenbosch University, Stellenbosch 7602, South Africa.
[4] Observatoire des Sciences et des Technologies, Centre Interuniversitaire de Recherche sur la Science et la Technologie, Université du Québec à Montréal, Montréal, QC, Canada.

*Correspondence to: vincent.lariviere@umontreal.ca

**Abstract:** Science is a cumulative activity, which can manifest itself through the act of citing. Citations are also central to research evaluation, thus creating incentives for researchers to cite their own work. Using a dataset containing more than 63 million articles and 51 million disambiguated authors, this paper examines the relative importance of self-citations and self-references in the scholarly communication landscape, their relationship with the age and gender of authors, as well as their effects on various research evaluation indicators. Results show that self-citations and self-references evolve in different directions throughout researchers' careers, and that men and older researchers are more likely to self-cite. Although self-citations have, on average, a small to moderate effect on author's citation rates, they highly inflate citations for a subset of researchers. Comparison of the abstracts of cited and citing papers to assess the relatedness of different types of citations shows that self-citations are more similar to each other than other types of citations, and therefore more relevant. However, researchers that self-reference more tend to include less relevant citations. The paper concludes with a discussion of the role of self-citations in scholarly communication.

**One-Sentence Summary:** This study provides evidence of career and gender effects in self-citations, and of a higher similarity of citing and cited papers in the case of self-citations than external citations.

**Main Text:**
Citation analysis has been used in research evaluation for almost five decades (*1-3*). What started as a tool to help researchers and librarians find relevant literature more efficiently (*4-6*) slowly became, after the creation Science Citation Index in 1963, a means to assess research and various levels, from individuals to institutions and countries (*7*). In this context, citation analysis has come under scrutiny from researchers across all disciplines. Several authors criticized bibliometrics and citation analysis for their limitations (*8*). Those limitations can be divided into those that relate to the coverage of the database (*9-11*), accuracy of citation data (*12*), adverse effects (*13-16*), overabundance of indicators (*17*), and citations being questionable indicators of research impact (*18-19*).

Literature on citation analysis has highlighted the diversity of roles of citations in scholarly papers. Based on papers published in high-energy physics, the classic study by Moravcsik and Murugesan (*20*) provides a typology of functions of citations, based on four non-exclusive dichotomies. Citations can be conceptual, related to theories or concepts contained in the cited document, or operational, based on methodological contributions of cited document. They can be organic, in which the cited document is necessary to understand the content of the citing document, or perfunctory, where the cited document is only peripheral to the citing document. Citation can also be evolutionary, in which citing document builds on the cited document, or juxtapositional, in cases where citing a document provides an alternative to the cited document. Finally, in cases where citing documents confirm the findings of a cited document, citations can be confirmative, or negative, when the citing and cited document disagree on findings or conclusions. While the presence of negative citations has often been considered to be a limitation of citation analysis (*18*), such negative citations account for a very small proportion of the citation record (*21-24*). Other issues related to the use of citations as indicators of research impact include uncitedness and uncited influences (*25-27*), coercive citations (*28-30*), and self-citations, which have been abundantly discussed in literature (*31-40*).

Self-citations occur where some features of the citing paper (author, journal, institution, country) are also found in the citing paper. Two conflicting hypotheses exist on the role of self-citations in scholarly communication. The first hypothesis is that self-citations are a normal feature of the cumulative nature of knowledge production (*33*), and that they don't change the ranking of researchers, institutions or journal in a drastic manner (*41-43*). As researchers' careers advance, they build on their previous work, and consequently cite it. In this context, self-citations are a normal feature of research activity—as well as an effect of the specialization of researchers in their research careers—and therefore could be considered in research evaluation. The second hypothesis relates to how researchers react to the use of citation analysis in research evaluation (*13, 44*), and to the incentives or institutionalized reward structures it creates for researchers to cite their own work. In this context, researchers would cite their work as often as possible in order to increase citation rates (*45*), with a large proportion of those self-citations perfunctory or irrelevant to the citing paper (*20*). In this context, self-citations' purpose is to increase one's scholarly impact, and should be excluded from research evaluations.

To examine these hypotheses, this paper performs two analyses. The first analysis examines 1) how self-citations and self-references evolve throughout the careers of researchers, 2) how they vary as a function of their research production, and 3) how their contribute to their h-index (*46, 47*). The distinction between self-citations and self-references—which will be described in the methods section below—is crucial here, as those two indicators behave in different manners throughout researchers' careers. Following previous research (*48-51*), we also assess how self-citations vary by gender. The second analysis, based on linguistic / thematic proximity between citing and cited papers, assesses whether self-citations are closer or more distant to the citing paper than non-self-citations. If self-citations are mostly forced or perfunctory, they should be more distant to the citing paper than non-self-citations or well-meaning, legitimate self-citations.

**Methods**
Many existing studies of self-citations suffer from methodological and terminological inaccuracies. The first issue is the distinction between self-citations and self-references, with self-citations being generally used and an umbrella term for both phenomena. While both indicators are based on the same act—the fact that an author cites a paper on which they are an author—the interpretation of self-citations and self-references is different. Self-referencing describes how much an author draws upon their own work to inform the present work. This requires having previous work, which is dependent upon seniority, production, and a focused research agenda. Self-citations, on the other hand, demonstrate the impact of the work upon the scientific community. A high self-citation rate would suggest that the work primarily serves to inform the author's own subsequent work and did not have high impact upon the rest of the scientific community. In mathematical terms, while both indicators are based on the same number in the numerator—which is the number of documents with an overlap between citing and cited authors—the denominator varies between references and citations. Self-referencing rates are calculated as the proportion of references made to other articles authored by the authors of the citing article(s). The self-citation rate is calculated as the proportion of citations received by a given article that come from documents on which he or she was an author. These are two very different types of metrics and associated concepts: if one wants to assess deviant behavior, self-references is the indicator to use, as the author has the control on both the numerator and the denominator, while a low self-citation rate may be due to having received a high number of citations—which can both be made by the authors and others—therefore increasing the denominator. Despite this distinction, we use the term self-citation as the generic term for both self-citations and self-references.

The second issue relates to the level at which self-citations are observed: author- or paper-level. For single-authored publications, author- and paper-level self-citations are the same. However, the vast majority of papers are the results of collaboration (*52-53*), and calculation of self-citations for collaboratively authored articles is more complicated. While a self-citation at the paper-level is observed when there is any overlap between the set of citing and cited authors, these self-citations may not apply in the same manner to the authors of the citing and cited paper. For example, in many cases, not all authors of the citing paper will also be authors of the cited paper and, therefore, one cannot attribute a self-citation or a self-reference to an author who is not active in the self-citation act. Such misattribution of paper-level self-citations to authors who are not authors of the cited paper, which significantly inflate self-citation rates, have been performed in a recent set of analyses (*54-55*). Fig. S1 presents a schematic relation of the difference between self-citations and self references, as well as between paper-level and author-level self-citations. While it makes sense to consider paper-level self-citations when analyzing their effect on the impact of scholarly papers, when one seeks to understand the self-citation practices of individuals, it makes no sense to consider references made by other authors of a paper as one's self-citations. As we are interested in researchers' behavior, therefore, self-citations and self-references are compiled at the researcher level.

An important challenge for compiling author-level self-citations is the availability of large-scale author disambiguation algorithms. This paper uses Clarivate Analytics' Web of Science, along with the author disambiguation algorithm developed at CWTS-Leiden (*56*). Publication data between 1980 to 2018 was used, for a total of 63,327,731 articles and 881,480,407 cited references to source items. Authors from those papers were disambiguated into 51,597,960 authors, using heuristics such as name and given name, institution, collaborators, cited references, topics and email addresses. The algorithm errs of the side of precision rather than recall, which means that it is more likely to split researchers into two entities rather than group two distinct researchers into one entity. We only included authors that have more than 5 publications in the analysis. As previously mentioned, self-citations and self-references were compiled at the individual researcher level, and year of first publication is used as proxy for career age (*57*). Each citation / reference was categorized into four types: 1) direct self-citations, where an author directly cites a paper on which they are a co-author; 2) co-authors self-citations: co-authors of the paper cite the paper; 3) collaborators citations: former collaborators cite the paper; and 4) external: other citations than the three above.

We weighted citations based on the year where the citation was made to control for citation inflation. We divided the total number of references made in each given year by the number of articles published in that year to get an average number of references by article for each year. We then normalized this average with this formula:

$$w_y = \frac{1}{\mu_{ref_y}/\max(\mu_{ref})},$$

Where $w_y$ is the weight of references made in articles published in year $y$, $\mu_{ref_y}$ is the average number of references made in each article published in year $y$, and $\max(\mu_{ref})$ is the maximum average references per publication across all years included in the analysis (1980 to 2019 were included, 2019 had the maximum value).

Of the 63,327,731 articles analyzed in the first part of the paper, 26,329,066 had an abstract and a disambiguated author; those were used for the citing/cited similarity analysis, for a total of 15,749,808 pairs of citing-cited articles. Since a given citing-cited pair can involve more than one author this leads to 1,689,855,285 citing-cited comparisons with unique authors. The text of each abstract was preprocessed to remove stop words and we took the stem (root) of every word. We then used a tf-idf representation of every abstract, and obtained the cosine similarity between each pair of citing-cited articles. We computed an average similarity score for each citation type of each author.

**Results**
Fig. 1 presents the percentage of self-citations and self-references as a function of career age, by domain. Trends are similar in all three domains, with percentages being generally higher in natural sciences and engineering, followed by social sciences, and then by health sciences. It shows that self-citations and self-references follow a totally different pattern as a function of career age. The percentage of self-citations start at the highest point early in the career, then decreases sharply until about the 10$^{th}$ year following the first publication, and then decreases at a slower pace, with the oldest academic age being the one where the percentage of self-citations being the lowest. This intuitively makes sense: early in the career, one's research papers are not known by many, and therefore, are more likely to be self-cited than cited by others. As career progresses, so does the external visibility of their research papers, which in turn is associated with a decrease in the percentage of self-citations. Co-authors' self-citations follow a similar pattern, although with lower extremes.

In contrast, the percentage of self-references starts at the lowest point, then sharply increases for the first five years, then increases at a slower rate for the rest of the career, with the highest percentage reached at the oldest academic age. This, also, can be explained by career progression: at the beginning of the career, authors do not have any previous publications to cite, and as they increase their number of papers, they have a higher number of previous papers to cite, therefore increasing their percentage of self-references. It is worth noting that, while self-citations are higher than self-references early in the career, the two curves shift at age 6-10, and the percentage of self-references become higher than the percentage of self-citations. Interestingly, co-authors references are relatively stable throughout the career, which is likely due to the fact that co-authors on papers can be of different academic ages. The percentage of both citations and references coming from collaborators increases from very low values at the beginning of the career to high values—especially for references—which is likely due to the fact that the network of previous collaborators increases over time. These patterns are also observed for different levels of research production and across all domains (Figs. S2-S5).

Direct and co-authors self-references are also more likely to be younger than those from former collaborators and other researchers (Fig. S6), which is likely a reflection of the faster integration of one's past work in their current work. Taken globally, this suggests that the self-citations and self-references are mostly driven by researchers' career progression and cumulative nature of research activities, and that this remains mostly invariant across domains.

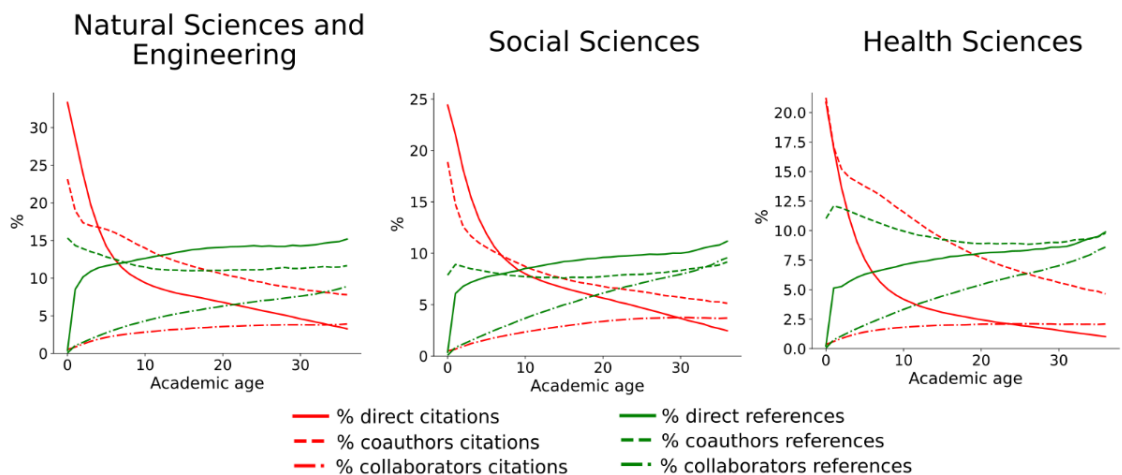

Fig. 1. Percentage of direct, co-authors, and collaborators self-citations and self-references, as a function of academic age, by domain

Fig. 2 presents, for each of the three domains, the cumulative effect of various forms of self-citations (direct, coauthors and collaborators) as a function of the observed h-index of researchers (see Fig. S10 for the effect of each type of citations individually). It shows that the relative importance of all types of self-citations rise sharply for low values of h-indices, and then stabilize quickly for direct and co-authors. The gain in the h-index obtained from direct self-citations hovers at about 3-5% in all three domains, irrespective of the h-index value. Logically, when direct self-citations are combined with co-authors self-citations, the gain in the h-index is higher but, again, remains stable between 8% and 15% across different h-index values. However, when collaborators' self-citations are included, the relative percentage of the h-index that is due to self-citations increases as a function of the h-index. For instance, while researchers in natural sciences with an h-index of 5 would see their index decrease by 15%, on average, if direct, co-authors, and collaborators self-citations were excluded, this percentage increases to 20% for h-indices of 30, and to almost 25% for h-indices of 60. This suggests clear individual gains that come, not only from the researchers themselves, but from those that are part of their research network (*58-59*). Although on average the impact of self-citations is low to moderate, some researchers would have half the h-index if not for self-citations (Fig. S11). This means that in the most extreme cases, some researchers with an h-index of 30 would have an h-index of 15 if self-citations at the level of articles were excluded.

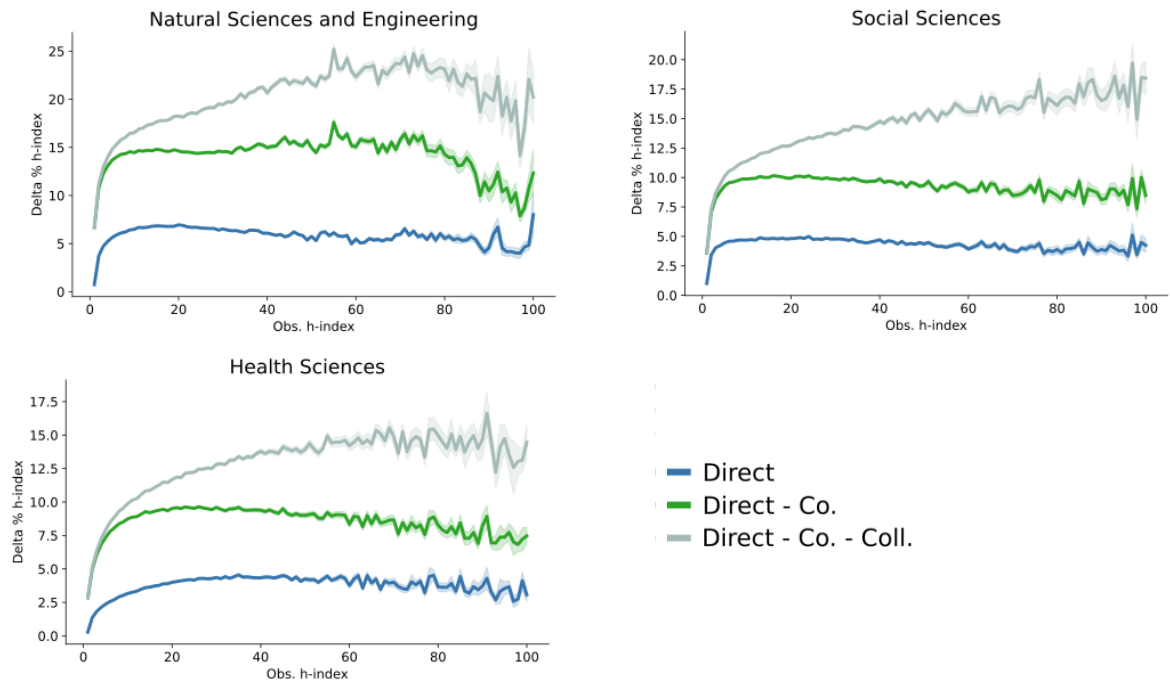

Fig. 2. Percentage increase of the of h-index that can be attributed to direct, co-authors and collaborators citations as a function of observed h-index, by domain.

Grouping researchers as a function of their percentile of self-references, we show that researchers' level of self-references is associated with external citations, gender, and academic age (Fig. S7). Researchers who exhibit high self-referencing behavior are also more cited by other researchers, controlling for their number of publications and discipline. Similarly, men are more likely to be in high percentiles of self-citers, even when controlling for productivity and impact levels. Although this is observed in every discipline, the trend is more pronounced in the social sciences. Older researchers in social sciences and health sciences are also more likely to be in high percentiles of self-citations (see also Fig. S8).

Assessment of the similarity between citing and cited abstracts shows that direct self-citations exhibit a higher level of similarity, followed by those from collaborators, and then by external citations (Fig. 3a,b). Those differences are also constant throughout various citation ages (Fig. 3c). However, gender differences are observed, with men's self-cited works being less similar than women's self-cited works, especially for direct self-citations (Fig. S9). This was particularly the case in the social sciences. This suggests proximity between citing and cited authors is associated with higher text similarity, which we can interpret as an indication that self-citations are more pertinent to the citing paper than external citations. However, when analyzing those trends as a function of researchers' percentage of self-references (Fig. 3d), we observe a sharp decline in the citing-cited article similarity as a function of researchers' percentage of self-references, which suggests that the more researchers' self-reference their own work, the less relevant those cited references are to the citing paper.

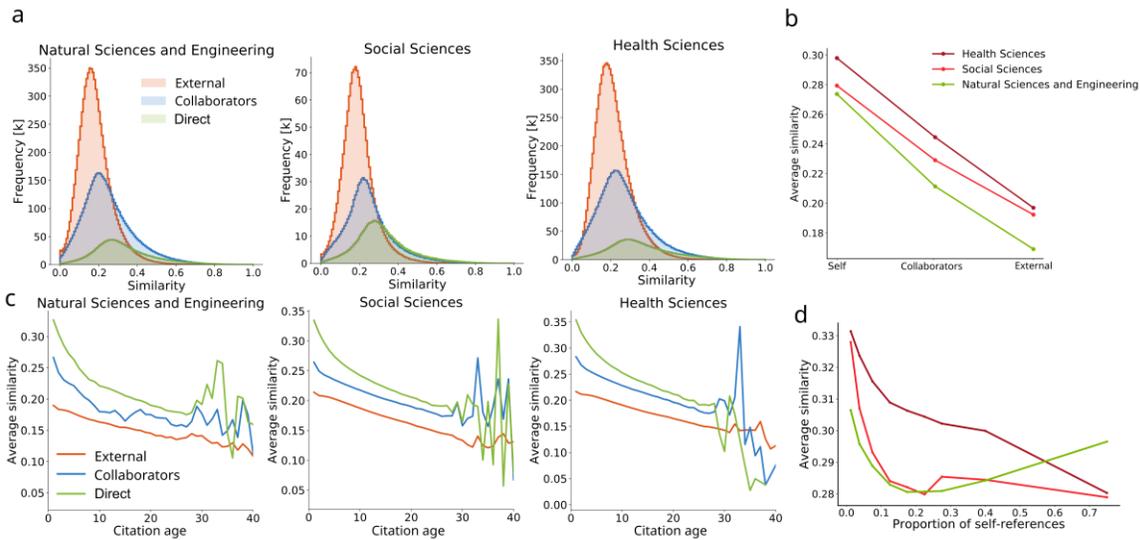

Fig. 3. a) Distribution of the similarity between citing and cited papers as a function of the type of citation, by discipline. b) average similarity of citation types, by discipline. c) Average similarity of citation types as a function of citation age, by discipline. d) average similarity of self-references, as a function of the percentage of self-references, by domain.

**Discussion and conclusion**
Citation-based indicators are at the heart of many national evaluation systems (*60-61*), university rankings (*62*), and tenure and promotion criteria (*63-64*). Although known to be imperfect indicators of research impact (*65*), there is general consensus they convey valuable information on how a piece of research influences the creation of new knowledge, and demonstrate some form of interaction between past and current research. Of the many limitations of citation analysis, self-citations are amongst the most discussed in the literature, with previous research focusing on self-citations at the journal (*66*), institution (*5*), paper (*67*), and author-levels (*68*). Critics argue that self-citations unduly inflate citation counts and are fundamentally different from external citations. It has been argued, therefore, that self-citations should be removed from citation indicators, especially when aiming at measuring scholarly impact. Citation inflation associated with self-citations are a particularly important problem in contexts where researchers are rewarded for performance on certain scientometric indicators by their institutions. Following Goodhart's law (*69*)—once a measure becomes a target, it ceases to become a reliable measure—the transparent application of citation-based indicators in research evaluation, and the deliberate inflation of citation counts, degrades the reliability of citations as accurate markers of credit for scientific ideas. Therefore, a better understanding of the extent to which researchers can influence their own citation rates is of utmost importance.

This paper makes five contributions to the literature on researchers' self-citations. First, it provides a methodological clarification of the difference between self-citations and self-references, as well as between paper- and author-level self-citations. Using large-scale data on disambiguated authors, it shows that aggregating paper-level self-citations at the researcher level (54-55) leads to misleading results, which largely overestimate researchers' self-citation and self-referencing practices. Second, it offers clear evidence that self-citations and self-references obey to different logics, and follow a different path over the course of a career. While the relative importance of self-citations decreases over time as researchers become more visible to colleagues, that of self-references increases quickly in the first couple of years of the career, then continues to increase at a slower pace. Those results can be explained by researchers' career trajectories and visibility: early in their careers, researchers' papers are less known—and therefore are more likely to be self-cited. As researchers progress in their careers, the external visibility of their papers increases, thus decreasing their percentage of self-citations. For self-references, the increase can be associated with the volume of previous papers that can be self-cited—which increases throughout researchers' careers—and to the fact that researchers mostly work on the same topics throughout their careers. These empirical results on the nature of self-citations and self-references also show that young documents are more likely to be self-cited, as it takes time for documents to be disseminated beyond their own authors. These 'mechanical' factors affecting self-citations suggest that researchers from different cohorts cannot be held to the same standards in terms of self-citations.

Third, at the individual researcher level, our results show that direct self-citations have, on average, little effect on the h-index, but can have a considerable impact for researchers that have high rates of self-citations. This suggests that, while self-citations are, in most cases, not affecting the reliability of citation analysis for assessing research impact, some researchers with unusually high levels of self-citations do. Fourth, it provides evidence that self-citations are correlated with external citations. That is, researchers who have high levels of self-citations are also highly cited by their colleagues. Women are also self-citing their work less than men. From a gender point of view, our results converge with previous research that reported that male scholars have a greater propensity to self-cite than female counterparts (*48*). These results are congruent with studies that show that women are relatively less prone to engage in self-promoting behavior at work (*70*), and that women tend to face relatively greater social penalties for self-promotion than men (*71*). In line with those, our results suggest that women require a greater threshold of relevancy in order to self-cite. This "higher bar" results in lower self-citation rates for women vis-à-vis men.

Finally, or results show that self-citations are more highly related to the citing paper than external citations, which suggest that, in most cases, self-citations are a normal feature of knowledge accumulation. However, such finding can be nuanced by the inverse relationship observed between text similarity and the extent of researchers' self-refencing practices: the higher researchers' percentage of self-references, the lower the similarity between citing and cited articles. This suggests that the subset of heavily self-citing authors is much more likely to exhibit questionable referencing behavior.

In sum, our large-scale analysis suggests that, in most cases, self-citations are an inherent feature of knowledge accumulation and of the progression of research careers. Our results show that, while there are self-citers who disproportionately cite their own work, self-citation is mostly a consequence of the cumulative nature of scientific work, of subject similarity within research programmes, and of the trajectory of scientific careers. All these factors make it highly likely—if not necessary—for self-citations to occur at the individual level.

**References and Notes**

1. Narin, F., & Carpenter, M. P. (1975). National publication and citation comparisons. Journal of the American Society for Information Science, 26(2), 80-93.

2. Narin, F. (1976). Evaluative bibliometrics: The use of publication and citation analysis in the evaluation of scientific activity (pp. 206-219). Cherry Hill, NJ: Computer Horizons.

3. Pinski, G., & Narin, F. (1976). Citation influence for journal aggregates of scientific publications: Theory, with application to the literature of physics. Information processing & management, 12(5), 297-312.

4. Garfield, E. (1955). Citation indexes for science. Science, 122(3159), 108-111.

5. Westbrook, J. H. (1960). Identifying Significant Research: Literature citation counting is evaluated as a means for identification of significant research. Science, 132(3435), 1229-1234.

6. Garfield, E., & Sher, I. H. (1963). New factors in the evaluation of scientific literature through citation indexing. American documentation, 14(3), 195-201.

7. Sugimoto, C. R., & Larivière, V. (2018). Measuring research: What everyone needs to know. Oxford University Press.

8. Garfield, E. (1979). Is citation analysis a legitimate evaluation tool?. Scientometrics, 1(4), 359-375.

9. Archambault, É., Vignola-Gagné, É., Côté, G., Larivière, V., & Gingras, Y. (2006). Benchmarking scientific output in the social sciences and humanities: The limits of existing databases. Scientometrics, 68(3), 329-342.

10. Larivière, V., Archambault, É., Gingras, Y., & Vignola?Gagné, É. (2006). The place of serials in referencing practices: Comparing natural sciences and engineering with social sciences and humanities. Journal of the American Society for Information Science and Technology, 57(8), 997-1004.

11. Mongeon, P., & Paul-Hus, A. (2016). The journal coverage of Web of Science and Scopus: a comparative analysis. Scientometrics, 106(1), 213-228.

12. van Eck, N. J., & Waltman, L. (2019). Accuracy of citation data in Web of Science and Scopus. arXiv preprint arXiv:1906.07011.

13. Seeber, M., Cattaneo, M., Meoli, M., & Malighetti, P. (2019). Self-citations as strategic response to the use of metrics for career decisions. Research Policy, 48(2), 478-491.

14. Baccini, A., De Nicolao, G., & Petrovich, E. (2019). Citation gaming induced by bibliometric evaluation: A country-level comparative analysis. PLoS One, 14(9), e0221212.

15. Quan, W., Chen, B., & Shu, F. (2017). Publish or impoverish: An investigation of the monetary reward system of science in China (1999-2016). Aslib Journal of Information Management, 69(5), 486-502.

16. Alperin, J. P., Nieves, C. M., Schimanski, L. A., Fischman, G. E., Niles, M. T., & McKiernan, E. C. (2019). How significant are the public dimensions of faculty work in review, promotion and tenure documents?. eLife, 8.

17. Wilsdon, J., Allen, L., Belfiore, E., Campbell, P., Curry, S., Hill, S., ... & Tinkler, J. (2015). The metric tide. Report of the independent review of the role of metrics in research assessment and management.

18. MacRoberts, M. H., & MacRoberts, B. R. (1989). Problems of citation analysis: A critical review. Journal of the American Society for information Science, 40(5), 342-349.

19. Gingras, Y. (2016). Bibliometrics and research evaluation: Uses and abuses. MIT Press.

20. Moravcsik, M.J. & Murugesan P. (1975). Some results on the function and quality of citations. Social Studies of Science, 5(1), 86-92.

21. Catalini, C., Lacetera, N., & Oettl, A. (2015). The incidence and role of negative citations in science. Proceedings of the National Academy of Sciences, 112(45), 13823-13826.

22. Baldi, S., & Hargens, L. (1995). Reassessing the N-rays reference network: The role of self citations and negative citations. Scientometrics, 34(2), 239-253.

23. Chubin, D. E., & Moitra, S. D. (1975). Content analysis of references: Adjunct or alternative to citation counting?. Social studies of science, 5(4), 423-441.

24. Lamers, W. S., Boyack, K., Larivière, V., Sugimoto, C. R., van Eck, N. J., Waltman, L., & Murray, D. (2021). Meta-Research: Investigating disagreement in the scientific literature. Elife, 10, e72737.

25. Nicolaisen, J., & Frandsen, T. F. (2019). Zero impact: A large-scale study of uncitedness. Scientometrics, 119(2), 1227-1254.

26. Hamilton, D. P. (1991). Who's uncited now?. Science, 251(4989), 25-25.

27. MacRoberts, M. H., & MacRoberts, B. R. (1988). Author motivation for not citing influences: A methodological note. Journal of the American Society for Information Science (1986-1998), 39(6), 432.

28. Wilhite, A. W., & Fong, E. A. (2012). Coercive citation in academic publishing. Science, 335(6068), 542-543.

29. Wren, J. D., & Georgescu, C. (2020). Detecting potential reference list manipulation within a citation network. BioRxiv.

30. Van Noorden, R. (2020). Signs of 'citation hacking' flagged in scientific papers. Nature, 584(7822), 508-509.).

31. Mishra, S., Fegley, B. D., Diesner, J., & Torvik, V. I. (2018). Self-citation is the hallmark of productive authors, of any gender. PLoS ONE, 13(9), e0195773.

32. Lawani, S. M. (1982). On the heterogeneity and classification of author self?citations. Journal of the American society for Information Science, 33(5), 281-284.

33. Glänzel, W., Debackere, K., Thijs, B., & Schubert, A. (2006). A concise review on the role of author self-citations in information science, bibliometrics and science policy. Scientometrics, 67(2), 263-277.

34. Glänzel, W., & Thijs, B. Schlemmer, B. (2004a). A bibliometric approach to the role of author self-citations in scientific communication. Scientometrics, 59(1), 63-77.

35. Glänzel, W., & Thijs, B. (2004b). The influence of author self-citations on bibliometric macro indicators. Scientometrics, 59(3), 281-310.

36. Glänzel, W., & Thijs, B. (2004c). Does co-authorship inflate the share of self-citations?. Scientometrics, 61(3), 395-404.

37. Chang, C. L., McAleer, M., & Oxley, L. (2013). Coercive journal self citations, impact factor, journal influence and article influence. Mathematics and computers in simulation, 93, 190-197.

38. Lariviere, V., Gong, K., & Sugimoto, C. R. (2018). Citations strength begins at home. Nature, 564(7735), S70-S71.

39. Costas, R., van Leeuwen, T., & Bordons, M. (2010). Self-citations at the meso and individual levels: effects of different calculation methods. Scientometrics, 82(3), 517-537.

40. Snyder, H., & Bonzi, S. (1998). Patterns of self-citation across disciplines (1980-1989). Journal of Information Science, 24(6), 431-435.

41. Medoff, M. H. (2006). The efficiency of self-citations in economics. Scientometrics, 69(1), 69-84.

42. Nederhof, A., Meijer, R., Moed, H., & Van Raan, A. (1993). Research performance indicators for university departments: A study of an agricultural university. Scientometrics, 27(2), 157-178.

43. Aksnes D.W. (2003), A macro study of self-citation. Scientometrics, 56 (2), 235-246.

44. Kacem, A., Flatt, J. W., & Mayr, P. (2020). Tracking self-citations in academic publishing. Scientometrics, 123(2), 1157-1165.

45. Fowler, J., & Aksnes, D. (2007). Does self-citation pay?. Scientometrics, 72(3), 427-437.

46. Hirsch, J. E. (2005). An index to quantify an individual's scientific research output. Proceedings of the National academy of Sciences, 102(46), 16569-16572.

47. Rad, A. E., Shahgholi, L., & Kallmes, D. (2012). Impact of self-citation on the H index in the field of academic radiology. Academic Radiology, 19(4), 455-457.

48. King, M. M., Bergstrom, C. T., Correll, S. J., Jacquet, J., & West, J. D. (2017). Men set their own cites high: Gender and self-citation across fields and over time. Socius, 3, 2378023117738903.

49. Deschacht, N., & Maes, B. (2017). Cross-cultural differences in self-promotion: A study of self?citations in management journals. Journal of Occupational and Organizational Psychology, 90(1), 77-94.

50. Pinheiro, H., Durning, M., & Campbell, D. (2022). Do Women Undertake Interdisciplinary Research More Than Men, and Do Self-citations Bias Observed Differences?. Quantitative Science Studies, 1-57.

51. Stiso, J., Oudyk, K., Bertolero, M. M., Zhou, D., Teich, E. G., Lydon-Staley, D. M., ... & Bassett, D. S. (2022). Modeling observed gender imbalances in academic citation practices. arXiv preprint arXiv:2204.12555.

52. Larivière, V., Gingras, Y., Sugimoto, C. R., & Tsou, A. (2015). Team size matters: Collaboration and scientific impact since 1900. Journal of the Association for Information Science and Technology, 66(7), 1323-1332.

53. Jones, B. F., Wuchty, S., & Uzzi, B. (2008). Multi-university research teams: Shifting impact, geography, and stratification in science. science, 322(5905), 1259-1262.

54. Ioannidis, J. P., Baas, J., Klavans, R., & Boyack, K. W. (2019). A standardized citation metrics author database annotated for scientific field. PLoS Biology, 17(8), e3000384.

55. Ioannidis, J. P., Boyack, K. W., & Baas, J. (2020). Updated science-wide author databases of standardized citation indicators. PLoS Biology, 18(10), e3000918.

56. Caron, E., & van Eck, N. J. (2014). Large scale author name disambiguation using rule-based scoring and clustering. In Proceedings of the 19th international conference on science and technology indicators (pp. 79-86). CWTS-Leiden University, Leiden.

57. Nane, G. F., Larivière, V., & Costas, R. (2017). Predicting the age of researchers using bibliometric data. Journal of Informetrics, 11(3), 713-729.

58. Milard, B., & Tanguy, L. (2018). Citations in scientific texts: do social relations matter?. Journal of the Association for Information Science and Technology, 69(11), 1380-1395.

59. Wallace, M. L., Larivière, V., & Gingras, Y. (2012). A small world of citations? The influence of collaboration networks on citation practices. PloS one, 7(3), e33339.

60. Debackere, K., & Glänzel, W. (2004). Using a bibliometric approach to support research policy making: The case of the Flemish BOF-key. Scientometrics, 59(2), 253-276.

61. Hicks, D. (2012). Performance-based university research funding systems. Research Policy, 41(2), 251-261.

62. Selten, F., Neylon, C., Huang, C. K., & Groth, P. (2020). A longitudinal analysis of university rankings. Quantitative Science Studies, 1(3), 1109-1135.

63. McKiernan, E. C., Schimanski, L. A., Nieves, C. M., Matthias, L., Niles, M. T., & Alperin, J. P. (2019). Meta-research: Use of the journal impact factor in academic review, promotion, and tenure evaluations. Elife, 8, e47338.

64. Shu, F., Quan, W., Chen, B., Qiu, J., Sugimoto, C. R., & Larivière, V. (2020). The role of Web of Science publications in China's tenure system. Scientometrics, 122(3), 1683-1695.

65. Bornmann, L., & Daniel, H. D. (2008). What do citation counts measure? A review of studies on citing behavior. Journal of Documentation, 64(1), 45-80.

66. Weiss, P. (1960). Knowledge: a Growth Process: Knowledge grows like organisms, with data serving as food to be assimilated, rather than merely stored. Science, 131(3415), 1716-1719.

67. Tagliacozzo, R. (1977). Self-citations in scientific literature. Journal of Documentation, 33(4), 251-265.

68. Kaplan, N. (1965). The norms of citation behavior: Prolegomena to the footnote. American documentation, 16(3), 179-184.

69. Goodhart, C. (1975). Problems of Monetary Management: The U.K. Experience. Papers in Monetary Economics. Vol. 1. Sydney: Reserve Bank of Australia.

70. Kay, K., & Shipman, C. (2014). The confidence code. The science and art of self. New York: Harper Collins.

71. Rudman, L. A., Moss-Racusin, C. A., Phelan, J. E., & Nauts, S. (2012). Status incongruity and backlash effects: Defending the gender hierarchy motivates prejudice against female leaders. Journal of experimental social psychology, 48(1), 165-179.

**Acknowledgments:** The authors would like to thank Stevan Harnad, Kyle Siler and Cassidy R. Sugimoto for comments on an earlier draft of this manuscript, as well as Richard Freeman and participants at the seminar in Economics of Science & Engineering at Harvard University; Jens Peter Andersen and participants at the Science Studies Colloquium at Aarhus University; and participants to the workshop on a New Synthesis for the Science of Science held at the Santa Fe Institute for their numerous comments and questions.

**Funding:** Funding for this analysis was provided by the Canada Research Chairs program, grant number 950-231768.

**Author contributions:**

    Conceptualization: PVL, VL

    Data Curation: PVL, VL

    Funding acquisition: PVL, VL

    Investigation: PVL, VL

    Methodology: PVL, VL

    Project administration: VL

    Resources: PVL, VL

    Supervision: VL

    Visualization: PVL

    Writing - Original Draft: PVL, VL

    Writing - Review & Editing: PVL, VL

**Competing interests:** Authors declare that they have no competing interests.

**Data and materials availability:** Restrictions apply to the availability of the bibliometric data, which is used under license from Thomson Reuters. Readers can contact Thomson Reuters at the following URL: http://thomsonreuters.com/en/products-services/scholarly-scientific-research/scholarly-search-and-discovery/web-of-science.html.

**Supplementary Materials**

Figs. S1 to S11

Citing paper

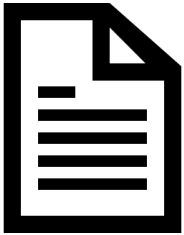

Larivière, V., & **Costas, R.** (2016). How many is too many? On the relationship between research productivity and impact. PLOS One, 11(9), e0162709.

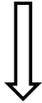

Cited paper

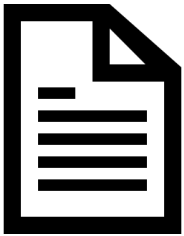

**Costas, R.**, van Leeuwen, T., & Bordons, M. (2010). Self-citations at the meso and individual levels: effects of different calculation methods. Scientometrics, 82(3), 517-537.

**Paper level**

- Self-citation / self-reference **for each citing and cited author**

**Author level**

- Self-reference for **R. Costas**
- Self-citation for **R. Costas**
- External reference for V**. Larivière**
- External citation for **T. van Leeuwen** and **M. Bordons**

Figure S1. Schematic representation of self-citations and self-references at the paper-level and at the author-level

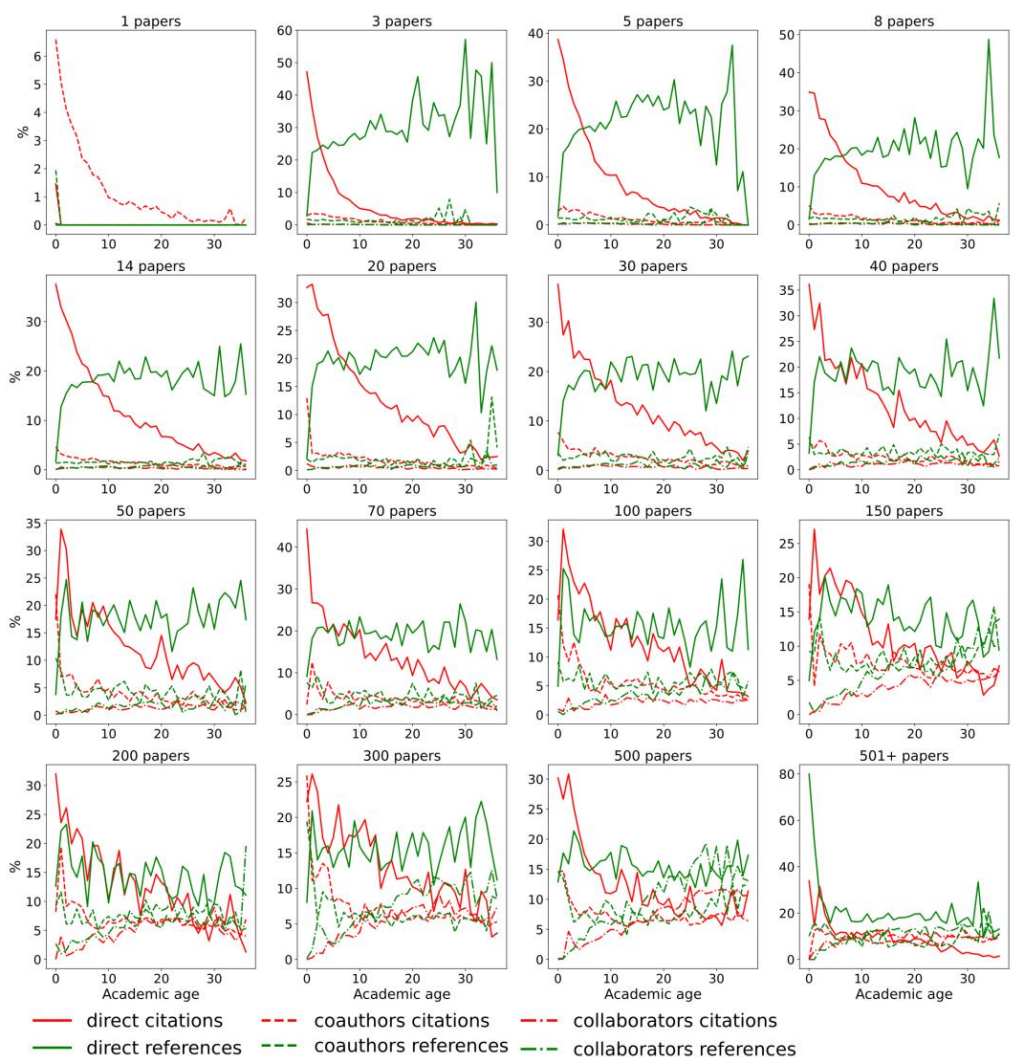

Fig. S2. Percentage of direct, co-authors, and collaborators self-citations and self-references, as a function of academic age and number of papers, arts and humanities.

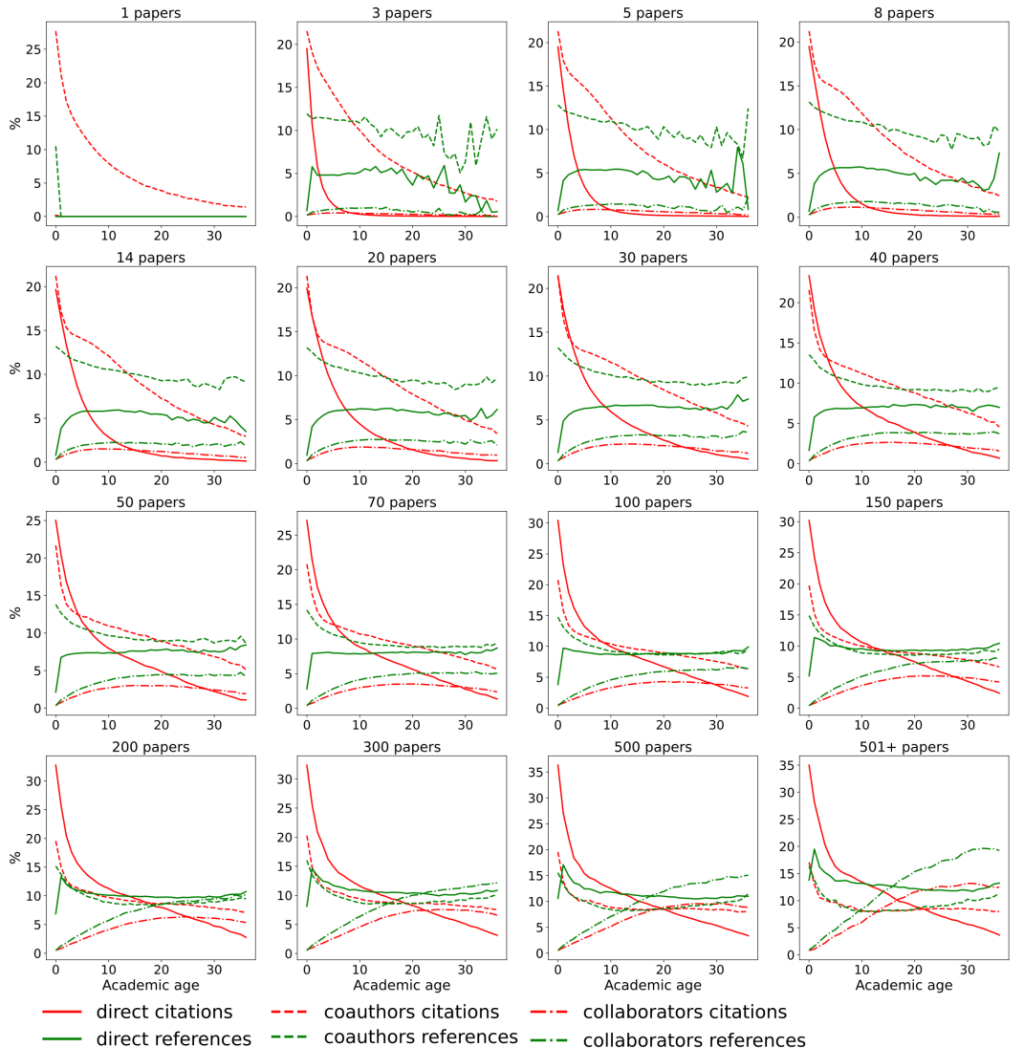

Fig. S3. Percentage of direct, co-authors, and collaborators self-citations and self-references, as a function of academic age and number of papers, health sciences.

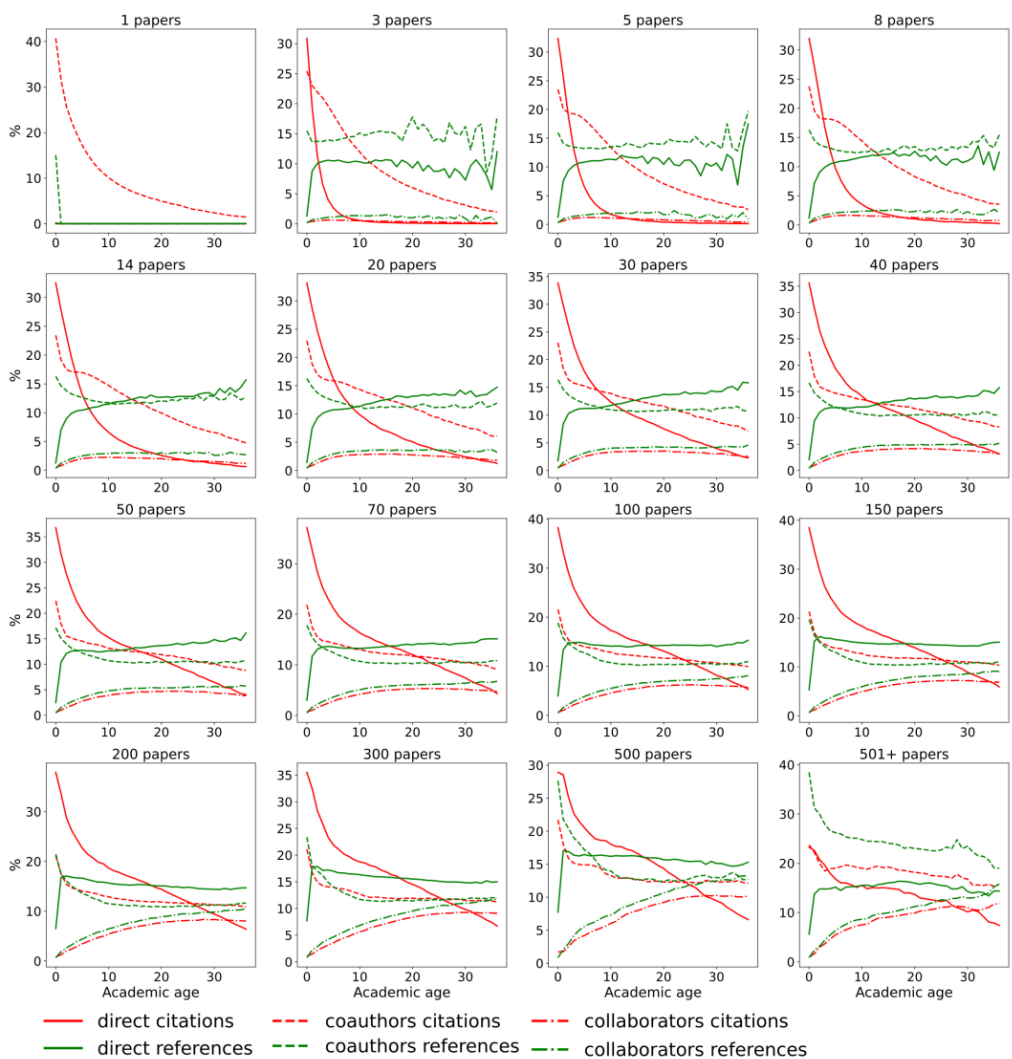

Fig. S4. Percentage of direct, co-authors, and collaborators self-citations and self-references, as a function of academic age and number of papers, natural sciences and engineering.

## Social Sciences

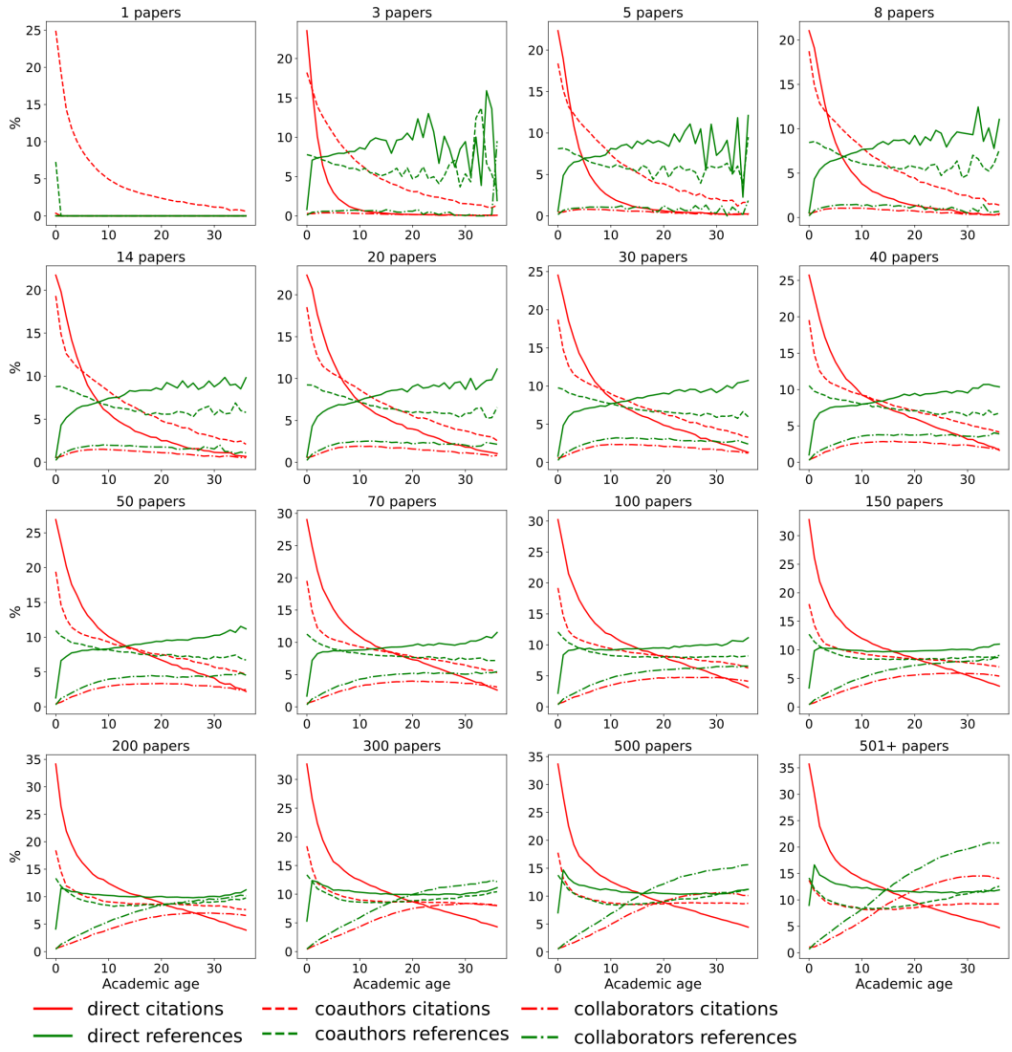

Fig. S5. Percentage of direct, co-authors, and collaborators self-citations and self-references, as a function of academic age and number of papers, natural sciences and engineering.

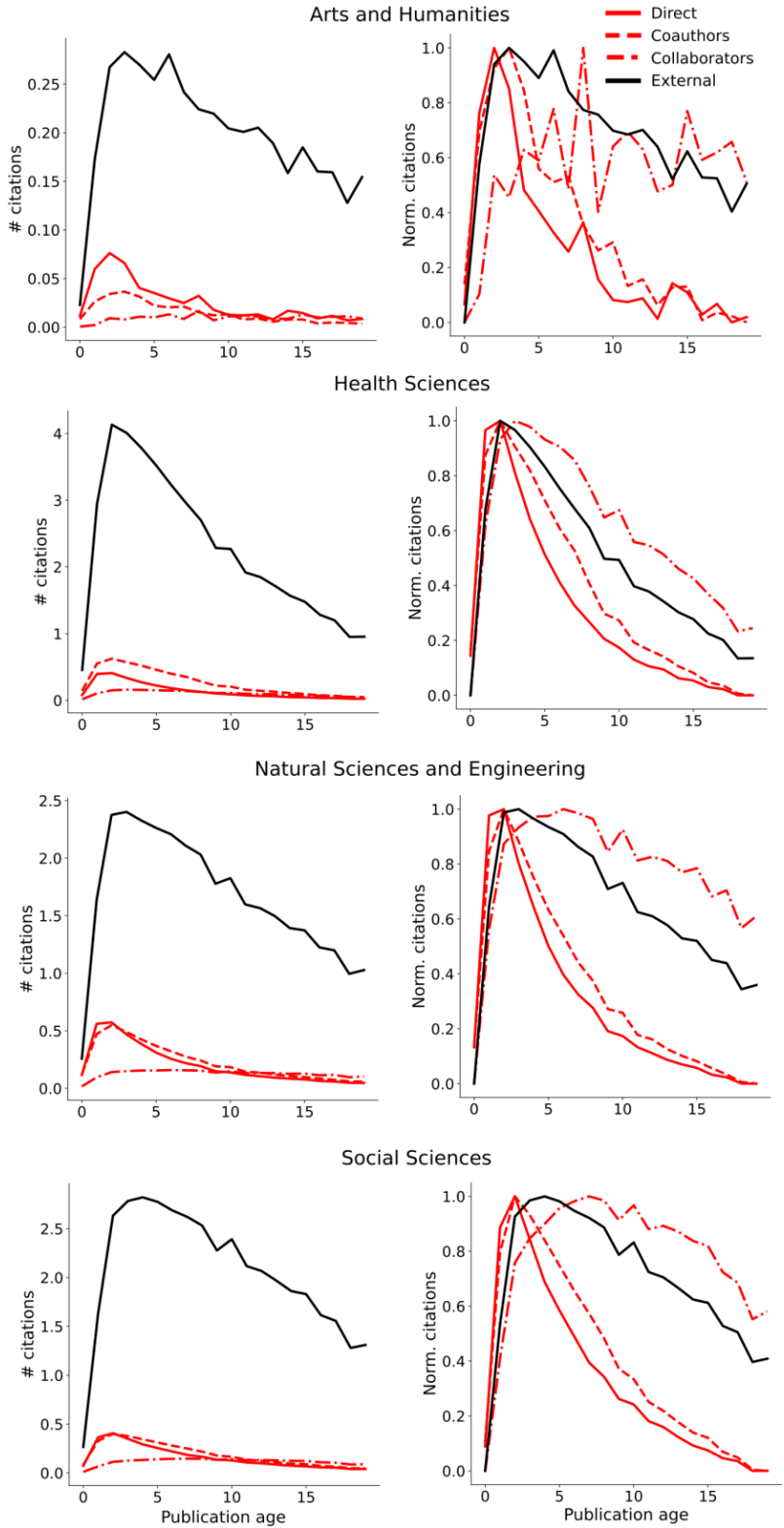

Figure S6. Age distribution of direct, co-authors, and collaborators self-references as a function of publication age, by domain. The left column shows the average number of citations per paper at different publication ages. The right column shows the citation types normalized with their own peaks in order to show their temporal dynamics.

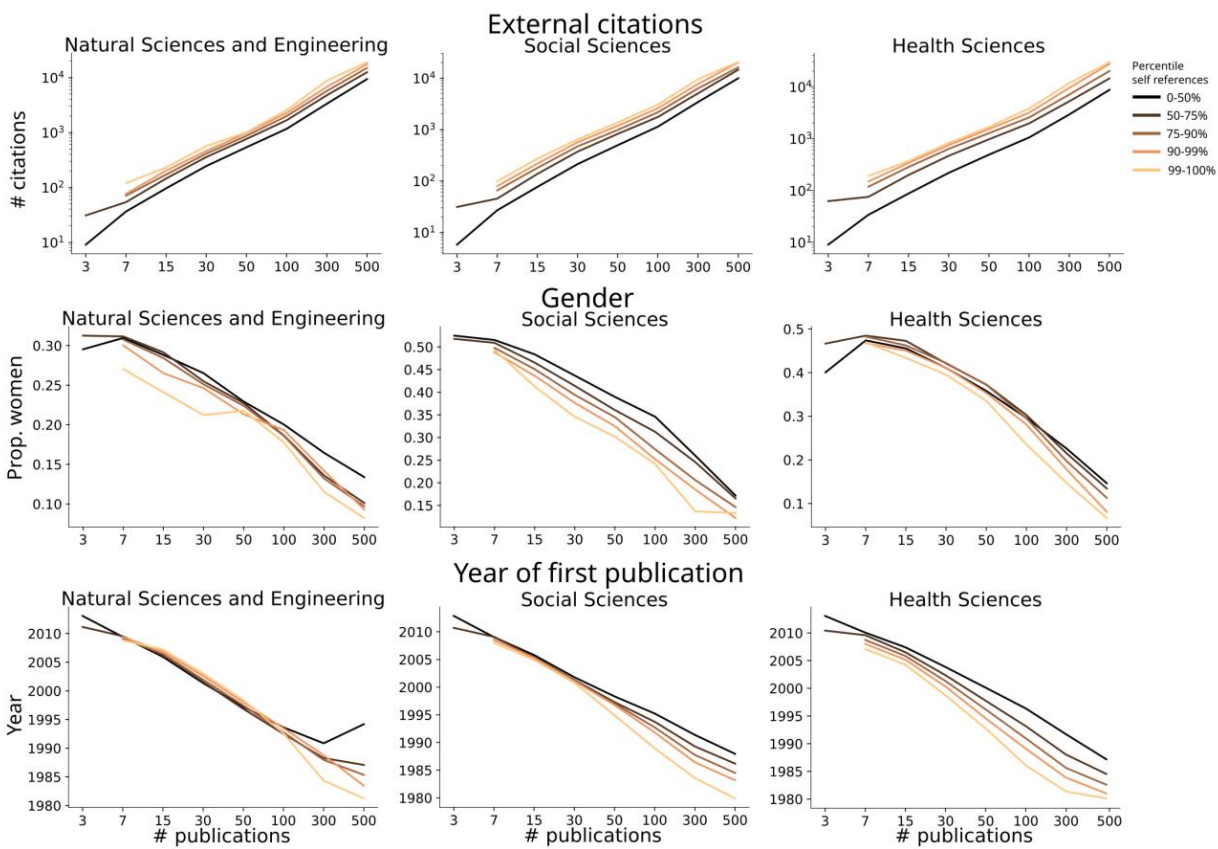

Fig. S7. Relationship between external citations (top row), percentage of women authorships (middle row) and year for first publication (bottom row) and number of publications, as a function of the percentile of self-references. Across all levels of publications activity (# of publications) and discipline, researchers with higher levels of self-references are also more cited than others. Researchers with a high proportion of self-references are more likely to be men, and to have started their academic career earlier.

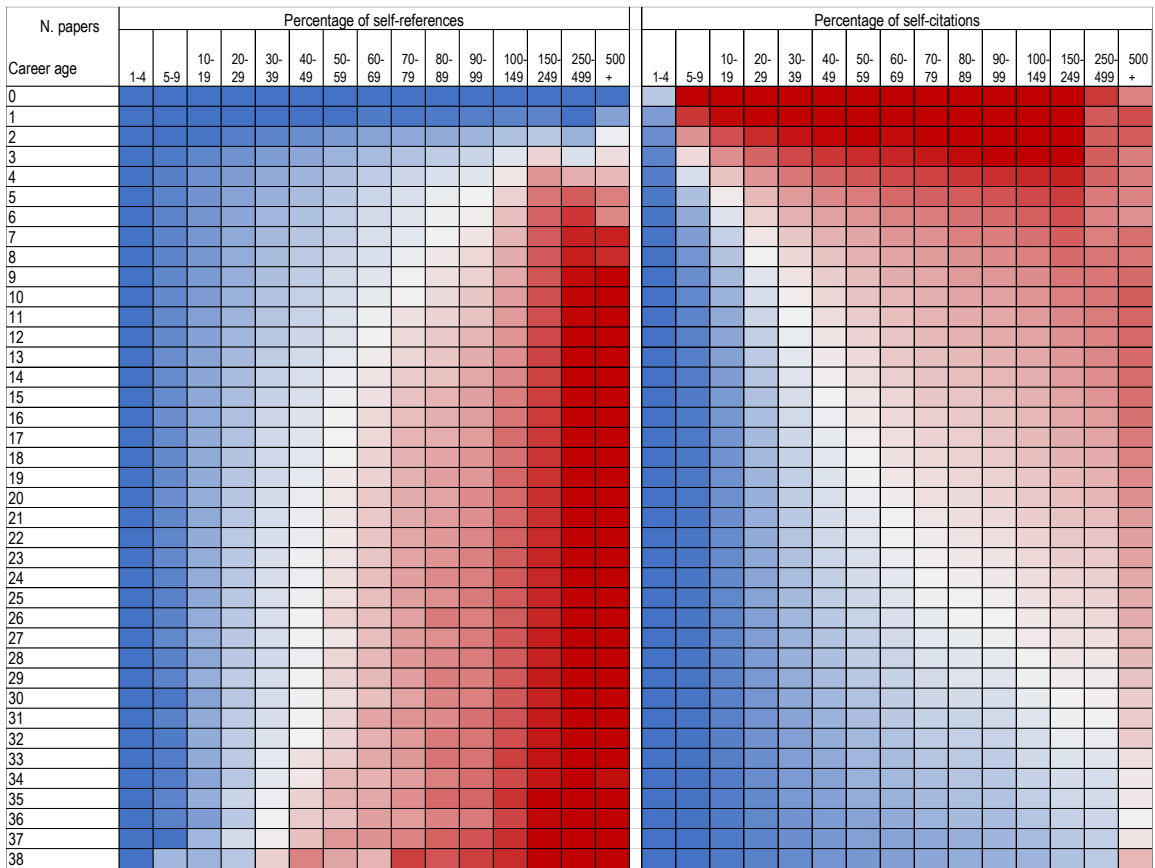

Figure S8. Percentage of self-citations and self-references, as a function of number of papers (x-axis) and of career age (y-axis). Red denotes higher percentages; blue denotes lower percentages.

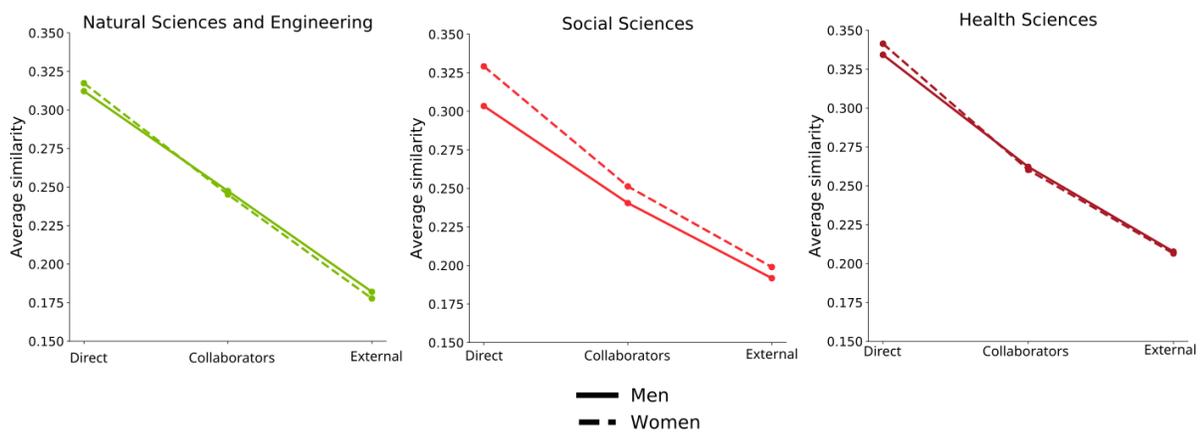

Fig. S9. Average similarly between citing and cited papers as a function of the type of citation, by gender and domain.

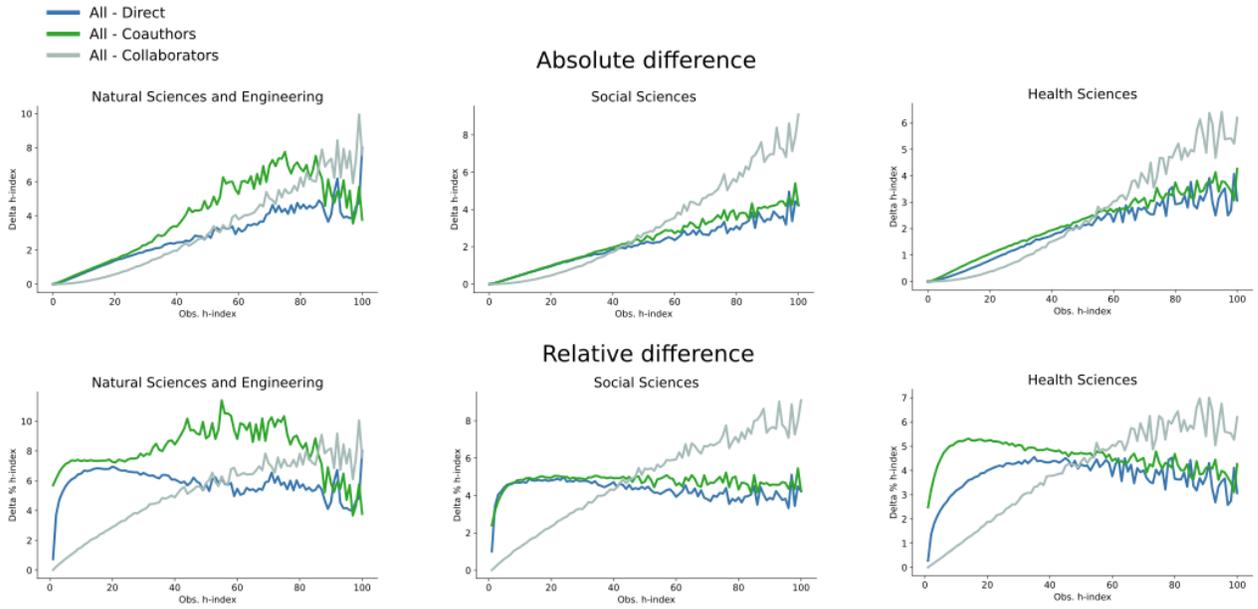

Fig. S10: Absolute and relative impact of each type citation individually (direct, coauthors and collaborators).

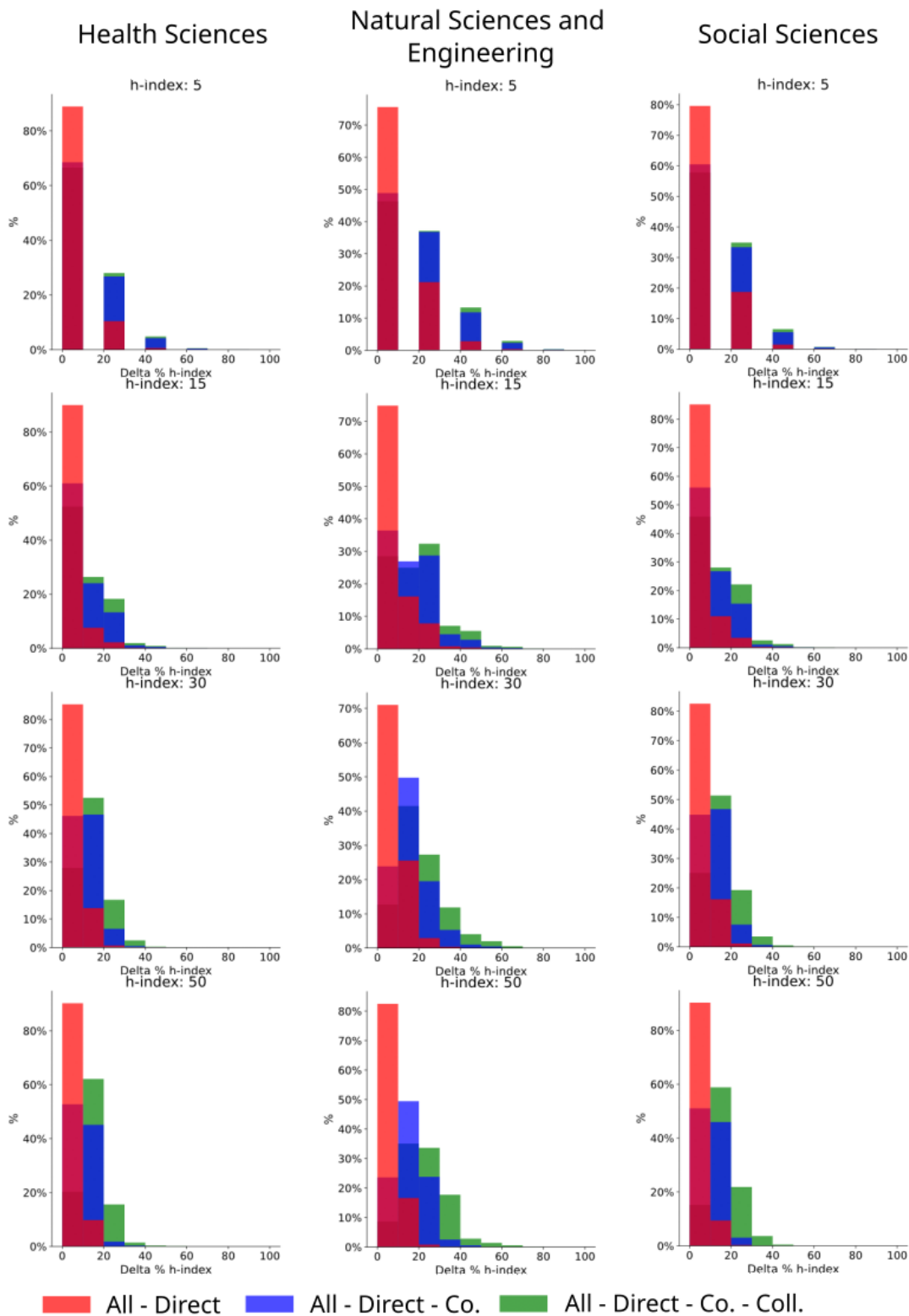

Fig. S11. Impact of citation types on the h-index. Disciplines are shown by column, and each row is a different observed h-index (5, 15, 30, 50). The red distribution shows the percentage of the observed h-index that can be explained by direct citations, the blue shows direct + coauthors and green the impact of direct + coauthors + collaborators.